\documentclass[preprint2]{aastex}

\def\aj{{AJ}}

\def\apj{{ApJ}}
\def\apjs{{ApJS}}

\def\deg{$^{\circ}$}

\def\dsec{{\rlap.}$^{\prime\prime}$}

\def\gax{{$\mathrel{\hbox{\rlap{\hbox{\lower4pt\hbox{$\sim$}}}\hbox{$>$}}}$}}

\def\lax{{$\mathrel{\hbox{\rlap{\hbox{\lower4pt\hbox{$\sim$}}}\hbox{$<$}}}$}}

\def\mnras{{MNRAS}}

\def\percm2{cm$^{-2}$}

\def\hii{\ion{H}{2}}

\received{}
\accepted{}
\slugcomment{Accepted for ApJ}
\shorttitle{Arp 299 Supernova Factory}
\shortauthors{Neff et al.}
\journalid{}{}
\articleid{}{}

\begin{document}

\title{A Supernova Factory in the Merger System Arp~299}

\author{Susan G. Neff}
\affil{NASA's Goddard Space Flight Center, Code 681, Greenbelt, MD 20771}
\email{neff@stars.gsfc.nasa.gov}
\author{James S. Ulvestad}
\affil{National Radio Astronomy Observatory, P.O. Box O, Socorro, NM 87801}
\email{julvesta@nrao.edu}
\author{Stacy H. Teng}
\affil{Department of Astronomy, University of Maryland, College Park, MD 
20742} 
\email{stacyt@astro.umd.edu}

\begin{abstract} 

We have imaged the nearby galaxy merger Arp~299 at arcsecond and
milliarcsecond resolution, using both the Very Large Array and 
the Very Long Baseline Array.  The large-scale radio emission from the 
merger contains 5 bright, compact radio sources embedded in diffuse 
emission, with diameters less than 200~pc.   
Supernova rates of 0.1 to 1 per year are required to 
produce the VLA-detected radio emission in these sources.
Two of the compact VLA radio sources, designated Source A and 
Source D, also have been detected and imaged at milliarcsecond 
scales.  Source A, which is associated with the nucleus of
one of the merging galaxies, contains five milliarcsecond-scale 
sources, each with a radio power between 100 and 1000 times that of the 
Galactic supernova remnant Cassiopeia~A.  Four of these have flat or 
inverted spectra and appear to be young supernovae.  Three of the 
VLBI-scale sources are located within 10~pc (projected) of one another,
and two are separated by less than 3~pc, indicating that they all 
may be within the same super starcluster or complex of such clusters. 
The brightest VLBI-scale source, A0, has an extremely
inverted spectrum, with $\alpha \gtrsim +2$ at gigahertz frequencies.
It seems to be the youngest supernova, which has not yet broken
out of its circumstellar shell.  The milliarcsecond radio sources 
within Source A appear to constitute a supernova factory, confirming
the presence of an extreme starburst that peaked at least a few
million years ago.

\end{abstract}

\keywords{galaxies: individual (Arp 299, NGC 3690) --- galaxies: interacting --- 
galaxies: starburst --- stars: clusters}

\vskip1000pt

\clearpage

\section{Introduction}
\label{sec:intro}

Arp~299 (NGC~3690, Mrk 171){\footnote{The Arp~299 merger is 
sometimes referred to as NGC~3690 (west) and IC~694 (east), although 
the designation of various merger components and nearby companions in 
the literature is inconsistent (see \citet{yam98} for a discussion); in
this paper we refer to the whole system as Arp~299.}}
is the ``original'' starburst galaxy (Gehrz, Sramek, 
\& Weedman 1983) and an obvious 
merger system that has been studied extensively at many wavelengths.
The galaxy distance is 41~Mpc for $H_0 = 75$~km~s$^{-1}$~Mpc$^{-1}$
\citep{tul88}, where 1\arcsec\ corresponds to 200~pc.
The merger in Arp~299 is similar in age to the ``Antennae''; 
both systems are considered to be in the ``early'' merger stage, slightly more
than one crossing time ($\geq 10^{8}$ yrs) before  
coalescence of the nuclei is expected
\citep{keel95,read98}.  An active starburst in Arp~299 is indicated by 
the high frequency of recent optically discovered supernovae in the galaxy, 
including SN 1992BU \citep{vb94}, SN 1993G \citep{for93}, 
SN 1998T \citep{li98,yam98}, and SN 1999D \citep{qiu99}.  
\citet{lai} used 2.2-$\mu$m adaptive optics imaging of Arp~299 and detected
what they believe to be many ``super star clusters'' (SSCs) 
that might include very high mass stars.  {\it Hubble Space Telescope\/} 
({\it HST}\/) FOC and NICMOS images \citep{meu, aah} 
reveal a population of young stellar clusters at ultraviolet, optical and
near-infrared wavelengths.

   At visible wavelengths, the disks of the interacting galaxies appear 
to overlap, but the individual nuclei still are separate ($\sim 4$~kpc) 
and distinguishable (See Figure 1).    Unlike the ``Antennae'', 
which has two clearly distinguished tidal tails, Arp~299 has a single
tidal tail, indicative of a prograde-retrograde interaction in the past
($\sim$700~Myr ago).  The most recently initiated interaction also appears
to be prograde-retrograde \citep{hibbard},  a spin geometry which drives 
most of the gas towards the galaxy center and maintains star formation 
at an elevated level for an extended time.   Most of the 
molecular gas in the system is concentrated near the nuclei, 
particularly the eastern (retrograde) nucleus \citep{sar91},
but CO is also found in other regions such as the bridge 
between the nuclei and the base of the tidal tails \citep{aalto,casoli}.    

   Arp~299's infrared luminosity ($ L_{IR} \geq 3 \times 10^{11} L_{\odot} $) 
is nearly in the Ultraluminous IR Galaxy (ULIRG) category \citep{hibbard}. 
The eastern nucleus, hereafter source A \citep{gehrz}, is the brightest 
component in the near-infrared, 
accounting for $\sim$50\% of the total infrared luminosity of the system 
\citep{aah}.  There are numerous other point-like near-infrared sources 
in Arp~299,  many of them in very dusty regions of the system \citep{aah}.  
Many, but not all, of the infrared objects correspond to sources detected
at radio wavelengths in VLA{\footnote{The Very Large Array (VLA), the Very
Long Baseline Array (VLBA), and the {\it Robert C. Byrd Green Bank Telescope} 
(GBT) are all operated by the National Radio Astronomy 
Observatory (NRAO), which is a facility of the National Science Foundation 
operated under cooperative agreement by Associated Universities, Inc.}}
images \citep{gehrz,zhao}.  Numerous \hii\ 
regions populate the system near star-forming regions and it appears that star
formation has been occurring at a high rate for about 10 Myr \citep{aah}.
   
The new and archival radio observations presented here are part of a 
larger project studying the star formation rate of mergers, to determine 
at which stage of the merging process the maximum star formation occurs. 
This paper focuses on the properties of the five strongest radio sources in 
Arp~299, investigating them on scales of $\sim$ 100~pc and of $\lesssim$1~pc.  
The population of weaker radio sources and the diffuse radio emission  
in the system will be the subject of a separate paper (Neff, Campion, and 
Ulvestad, in prep).

\section{VLA Observations, Archival Data, and Imaging}
\label{sec:vla}
New observations of Arp~299 were made using the VLA
\citep{vla} at 4.9 and 8.4~GHz;  these wavelengths were chosen because they 
are the most sensitive bands on the VLA.  The new 4.9 and 8.4~GHz
observations were made on 2000 October 24, 2001 March 19, and 2002 April 29.
In 2000 October, the system was observed at 4.9~GHz with the {\bf A} 
configuration.  Eight-minute integrations of the galaxy were alternated 
with 1-minute scans of the phase calibrator J1128+5925 over a period of four
hours; two 2-minute observations of 3C 286 were used for amplitude calibration.  
Arp~299 was observed at 8.4~GHz in the {\bf B} configuration in 2001 March, with 
a 
similar observing strategy, except that the total observation time of four 
hours was spread out over an 8-hr period.  

There are many additional observations of Arp~299 
in the VLA archives, mostly from supernova monitoring campaigns.  We selected a 
combination of data sets covering a wide range of baselines to complement the 
new 4.9 and 8.4~GHz observations and to maximize the aperture-plane coverage 
of the resulting combined data.  We used only observations with 
more than 200 seconds of integration on Arp~299.  All of the selected 
observations had a bandwidth of 50 MHz in each of two intermediate 
frequency channels for both right and left
circular polarization.  Each observation used 3C 286 as the flux 
calibrator and was calibrated to the primary flux density scale of 
\citet{baa77}, as modified very 
slightly by VLA staff to account for source variability.  
We obtained 14 sets of archival data at 8.4~GHz and 11 sets of archival
data at 4.9~GHz.  We also retrieved 
one observation at 1.4~GHz.  
Observations exist in the VLA archive at higher
frequencies (15 and 22 GHz); however we did not use these 
because the calibration
is less certain and the data give inconsistent results.  Details of all new 
and archival observations used in this paper 
are listed in Table~\ref{tab:vlaobs}.  
The noise levels for each individual image are listed; these images
were made with the ``robust'' parameter for data weighting 
that provides the best combination of high sensitivity and
high resolution \citep{bri95,brig98}.
The combination of the archival data with the new observations 
provided very high sensitivity over a wide range of spatial scales
at 4.9 and 8.4~GHz.  Although the 
image at 1.4~GHz is not of comparable sensitivity 
to the 4.9 and 8.4~GHz images, it is adequate for comparisons of 
emission from the strongest compact radio sources.

Each individual data set was calibrated, flagged, imaged, self-calibrated,
and re-imaged using standard image processing techniques 
in the Astronomical Imaging Processing System, AIPS \citep{aips}. 
At 4.9 and 8.4~GHz, the multiple data sets at each wavelength were 
combined into a single data set and another cycle of imaging plus
self-calibration and imaging produced images with high resolution 
and low noise.  The data processing included the removal of two 
background radio sources within the field of view.  They are a companion
E/S0 galaxy located 1\arcmin\ to the northwest (sometimes referred to as 
IC~694 \citep{yam98}), and a strong background
radio galaxy $\sim 1.5$\arcmin\ southeast of Arp~299 
(centered approximately at J2000 right ascension $11^h 28^m 29^s$,  
declination 58\arcdeg 32\arcmin 05\arcsec).

The default 8.4~GHz combined image was of higher resolution than the 4.9~GHz 
image.  
Thus the data weighting for the 8.4~GHz data was adjusted so that 
the resulting 8.4~GHz image matched the 0\dsec 38 $\times$0\dsec 31 resolution 
of the 4.9~GHz image; this adjustment increased the rms noise in the 
combined 8.4~GHz image.    Figures~\ref{fig:6cm} and 
\ref{fig:4cm} show the combined images of Arp~299 at 4.9 and 8.4~GHz, 
respectively.
The 1.4~GHz image (Figures~\ref{fig:20cm}) could not be adjusted to 
match the resolution of the 4.9 and 8.4~GHz images,
so additional images at 4.9 and 8.4~GHz were 
made to match the resolution of the 1.4~GHz image.   These images 
are not shown, but are discussed occasionally below and are used in
some of the analysis.

All of the images show five dominant compact radio sources.  The four 
strongest radio sources (labeled A, B1, C, and C$'$) have been identified 
previously by \citet{gehrz}.  The fifth brightest compact radio source, 
which we have labeled D after the \citet{gehrz} notation for the region 
where this source is located, was identified previously at radio wavelengths 
by \citet{hua90} and discussed by \citet{zhao}.   Source D clearly is 
visible in the 1.6-$\mu$m NICMOS image shown by \citet{aah}, and in
the HST 814 nm image (Figure 1).   
In our new radio images, Source B has been resolved into two sources, B1 
and B2, for the first time at
radio wavelengths (see Figures~\ref{fig:6cm} and \ref{fig:4cm}); 
these sources clearly are separated in near-infrared and optical images
shown by \citet{lai} and \citet{aah}. 

 Sources A, B1, C, C$'$, and D are the focus of this paper. 

\section{VLBA Observations and Imaging}
\label{sec:vlba}

We observed Arp 299 using Very Long Baseline Interferometry (VLBI)   
to search for possible active galactic nuclei (AGNs) or radio-luminous 
supernovae (SNe) within Sources A, B1, C, C$'$, and D.

Three sets of VLBI observations were obtained in 2002 and 2003.
In all cases, dual polarizations were observed, and Arp~299 
was phase-referenced 
\citep{bea95} to the calibration source J1128+5925.
The calibrator is only 0.9\deg\ away from the galaxy, allowing for 
good atmospheric calibration.  
On 2002 April 29, we observed the galaxy at 2.3~GHz using the 
VLBA$^{2}$, and including the 100 by 110 meter GBT$^{2}$
for increased sensitivity.  The observation 
was made with 32~MHz total bandwidth and 2-bit sampling, for
a total recorded data rate of 256~Mbit~s$^{-1}$.  
One-minute scans of the calibrator and 2.5-minute scans of the 
galaxy were alternated; total integration time on the galaxy 
was approximately 5 hours.  A second 
observation was made on 2003 February 9 using only the VLBA.  The galaxy 
was observed at 8.4~GHz in the same observing mode as in 2002; 
1-minute scans on the calibrator and 3-minute scans 
of the galaxy were alternated.  The total time on the galaxy was 
approximately 7 hours.  Finally, a third follow-up (``target 
of opportunity'') observation was made on 2003 April 30/May 1 
(hereafter 2003 May 1), using only the VLBA, at 2.3, 5.0, 8.4, and 15~GHz
These observations used only 16~MHz total bandwidth and
half the data rate of the previous observations.
Galaxy scan times ranged 
from 1 to 3 minutes in the phase-referencing.  The integration time at
each of the four wavelengths was selected to achieve equal noise levels at all
bands;
in the final datasets, the noises at 2.3 and 15~GHz were slightly higher 
due to ambiguous phase connection (15~GHz) and radio-frequency
interference (2.3~GHz). 
Details of the VLBA observations are listed in Table~\ref{tab:vlbaobs}.
Here, the listed noise levels are those for which ``natural'' weighting
was used; this maximizes sensitivity, at some expense in resolution
(Briggs et al., 1998).

Calibration, imaging and self-calibration of the VLBA data also were carried out 
using AIPS.  Amplitude calibration was based on system temperature 
measurements made during the observations, and has an accuracy of $\sim 5$\%. 
Instrumental delays, delay rates, and phases were 
calibrated by using the phase-reference calibrator and then applied to the 
galaxy.  Ionospheric corrections also were made using data provided by the 
Jet Propulsion Laboratory{\footnote{IONEX files can be obtained by FTP from 
cddisa.gsfc.nasa.gov.}}.   The data were processed
using short correlator integration times (0.5~seconds) and enough spectral
channels (32) to enable imaging at all locations in the merger without
losses in peak intensity due to bandwidth or time-average smearing
\citep{bri98}.  Self-calibration was carried out when possible in our 
low signal/noise data, and was most successful for the first epoch 2.3~GHz
observations, due to the highly sensitive baselines to the GBT.

The 2.3~GHz image of Source~A from 2002 (Figure~\ref{fig:vlba13cm-wide}) 
contains four unresolved compact sources located within
0\farcs15 of the VLA position for this source. 
A fifth unresolved object is detected at the position of Source D.  The other 
relatively strong VLA sources, B1, C, and C$'$, were not detected.
The initial 8.4~GHz image from 2003 February showed all the sources
present in 2002 - four in Source A and one in Source D.  Efforts 
to self-calibrate the data at quite low
signal/noise caused some of these sources to disappear, while others 
appeared in the images.  Therefore, we use the original 
phase-referenced images in this paper.  
As in 2002, B1, C, and C$'$ remained 
undetected.  However, a surprising result, shown in 
Figure~\ref{fig:vlba13cm-narr}, was the appearance of a new
$\sim 3$~mJy source, dubbed A0 (as the strongest source),
only 13~mas (2.6~pc) from A1.  This source is discussed in more 
detail in Section~\ref{sec:sourceA}.

Since we could not use self-calibration on Arp 299 in 2003 
February, we employed the check source for the phase
referencing, J1127+5650 (located roughly 2.5\deg\ from J1128+5925),
to estimate the coherence loss due to imperfect phase calibration.  Comparing
the original and self-calibrated images for this check source, we find a 
coherence loss of 24\%.  The phase connection for the
reference source is good, with apparent changes (depending on
baseline length) of $20^\circ$ to $50^\circ$ between successive
observations of the reference source; linear interpolation between
these scans probably gives a phase error with respect to time of
no more than $\sim 10^\circ$, which could account for only a small 
fraction of a 24\% coherence loss.  Therefore, we infer that
most of the coherence loss is caused by spatial structures in
the atmosphere, in which case the phase error is likely to scale
with the angular separation of reference source and target.
Since Arp~299 is only 0.86\deg\ from J1128+2925, we infer that
the loss of coherence for the galaxy should have been $\sim 1/3$
that for the check source, or about 8\%.  Therefore, we have increased
the measured source flux densities at this epoch by 8\%; since the
uncertainty in this estimate of coherence loss may be as much as a
factor of two, we also have added 8\% (in quadrature) to the other
error sources in computing the final flux density errors.

The appearance of A0 motivated a target-of-opportunity proposal
for multi-frequency VLBA imaging, in order to determine the spectrum
of this new source.  These observations in 2003 May 
were carried out with the VLBA alone, used lower bandwidth than the
previous observations, and did
not include the GBT; therefore they have higher rms noises
than the earlier observations.
Source A0 was not detected above the $2\sigma$ level at 
2.3~GHz in May 2003.  Components A1, A2, A3, and A4 
are detected at levels consistent with their previous sizes and flux densities, 
given the much higher error levels.  Source D changed significantly between
2002 April and 2003 May: the 2.3~GHz flux density in 2003 was more than
6 times higher than it was a year earlier, and Source D also became
resolved at both 8 and 15~GHz (Figure~\ref{fig:vlba-d}).

The properties and characteristics of these compact sources are 
discussed below. 

\section{Properties of the Compact Radio Sources}
\label{sec:sources}

\subsection{VLA Results}

In both the high-resolution images and those matched to the 1.4~GHz resolution,
single Gaussian models were fitted to the five strongest 
compact radio sources.  If the model fitting 
indicated that the compact source was consistent with being unresolved, 
the source was re-fitted using a point-source model. 
Results of these fits are given in Tables~\ref{tab:vlafits} and 
\ref{tab:matched}.
Flux density errors combine the noise, the fitting
errors, and possible systematic calibration offsets, added in quadrature.
For well-calibrated data, the overall amplitude scaling relative to
\citet{baa77} is reported by the VLA Calibrator Manual{\footnote{The latest
version of the VLA Calibrator Manual is accessible at 
http://www.vla.nrao.edu/astro/.}}
to be 1\% to 2\% at 1.4, 4.9, and 8.4~GHz. 
However, because most of our data are
archival, with a variety of different weather qualities and observing
strategies, we have conservatively assumed flux-scale errors of 5\%.

Peak brightness temperatures at 8.4~GHz were calculated for Sources A, B1, 
C, C$'$, and D using the flux densities and sizes found at the highest
resolution (see Table~\ref{tab:vlafits}), and are given in 
Table~\ref{tab:vlaprop}.
The brightness temperatures $T_{\rm B}$ were calculated using
Equation (1) from \citet{falcke}, modified for an elliptical gaussian:  
\begin{equation}
T_{\rm B}= 1.8\times10^3 \biggl(\frac{S_{\nu}}{\mathrm{mJy}}\biggr) 
\biggl(\frac{\nu}{\mathrm{GHz}}\biggr)^{-2} \biggl({\theta_1 
\theta_2\over {\rm arcsec}^2}\biggr)^{-1} \mathrm{K}\ .
\label{eqno1}
\end{equation}
Here, $S_\nu$ is the flux density at frequency $\nu$, with $\theta_1$ and
$\theta_2$ being the fitted full widths at half maximum 
of the major and minor axes of the sources.
Total monochromatic radio powers for each source were derived from the
matched-resolution images (see Table~\ref{tab:matched}).  These powers, given
in Table~\ref{tab:vlaprop}, typically are 
thousands of times the power emitted by the Galactic supernova remnant
Cassiopeia~A, which is assumed to have a present 8.4-GHz flux density of
460~Jy (derived from Baars et al. 1977) at a distance of 3.4~kpc
\citep{ree95}.  The inferred brightness temperatures and radio powers 
are as expected from complexes of hundreds to thousands of supernova
remnants, although we note that a few individual supernova remnants in dense
environments have been observed to show comparable luminosities
\citep{wei90,vand93}. 

Figure~\ref{fig:lowres} shows the radio spectra of the five 
compact components listed in Table~\ref{tab:matched}.  
On VLA scales, Sources A, B1, C, and C$'$ all have spectra that 
decrease slightly with increasing frequency, while
VLA source D has a spectrum that appears to peak at frequencies above
8~GHz. 
Two-point  spectral indices, $\alpha$ ($S_{\nu}\propto\nu^{+\alpha}$),
where $S_{\nu}$ is the flux density at frequency $\nu$), 
were calculated between 4.9 and 8.4~GHz from the images made with the 
same 1\farcs21$\times$0\farcs88 beam, and are listed in 
Table~\ref{tab:vlaprop} (Supernova rates in the table are derived
in section 5.3, below).
In spite of the small frequency range spanned, errors in 
these spectral indices are fairly small, $\sim 0.13$, because the 
five sources under study are detected with high 
signal/noise ratios.  
The spectral indices of $-0.5$ to $-0.65$ for Sources A, B1, C, and C$'$
are consistent with collections of typical galactic supernova remnants 
\citep{wei02}.
The spectrum of Source D is more enigmatic; we return to it in 
Section~\ref{sec:sourceD}.

\subsection{VLBA Results}

In all VLBA images, compact sources were visually identified and then fitted 
to a gaussian using the AIPS task JMFIT.  Details of the fits and flux 
measurements of the VLBA-scale compact sources are listed in 
Table~\ref{tab:vlbafits} and Table~\ref{tab:a0vlbi}.
Random errors are assigned as the quadratic sum of the image noise levels
(Table~\ref{tab:vlbaobs})
and the amplitude calibration errors, estimated to be 5\% at most 
wavelengths and (conservatively) as 10\% at 15~GHz.  In addition, for data
that could not be self-calibrated in 2003 February, the flux density at
8.4~GHz has been increased by 8\%, with an 8\% error added in quadrature,
as inferred from observations of the check source J1127+5650
(see Section~\ref{sec:vlba}). 

An important result of this work is the VLBI identification of
a number of compact radio sources with high brightness temperatures, 
embedded within Source~A and in Source~D.  Table~\ref{tab:vlba-tb}
gives the brightness temperatures of these milliarcsecond-scale
components, measured in the highest sensitivity 8.4~GHz VLBI images
from 2003 February.  For the unresolved sources, size upper limits
for each dimension are taken conservatively to be half the diameter 
of the half-power beamwidth.  In addition to the brightness temperatures,
powers relative to the galactic supernova remnant Cassiopeia~A (Cas~A) 
are listed, as are monochromatic radio powers at 8.4~GHz.  
Source D is assumed to be
at the distance of Arp~299 (see Section~\ref{sec:sourceD}); 
two values are given for
this object, one when it was unresolved in 2003 February, and one
when it was resolved in 2003 May.

The VLBI-scale sources have very different spectra, as can be seen
in Table~\ref{tab:a0vlbi} and in Figure~\ref{fig:vlbaspec}, which 
show the individual
source flux measurements  at four frequencies in 2003 May. 
Source A0 has an extremely inverted spectrum:  in May 2003 the 
5.0 and 8.4~GHz flux densities implied 
a spectral index of $\alpha>+1.76\pm 0.37$.  Comparison between 
the (non-contemporaneous) flux measurements of a $3\sigma$ upper 
limit of 0.11~mJy at 2.3~GHz in 2002 April, and   
the 8.4~GHz value of $3.23\pm 0.28$~mJy in 2003 February, yields 
$\alpha \gtrsim +2.6$ for Source A0, which declines more steeply at
long wavelengths than is expected even for
synchrotron self-absorption.   Source A1 has a non-thermal spectrum,
with $\alpha =-1.25\pm 0.28$ between 2.3 and 8.4 GHz.  Sources A2, A3, 
and A4 appear to have 
fairly flat spectra although the S/N in these (short) measurements
is not impressive.  However, if we assume no variability at 2.3
and 8.4 GHz between 
2002 April and 2003 February, all three sources have flat or slightly
inverted spectra.  In May 2003, Source D had a relatively flat spectrum,
rising somewhat at the higher frequencies.
 Although we do not know definitively 
what spectrum VLBI Source D had earlier, we do know that its 2.3~GHz
flux density increased from 0.30~mJy in Feb 2002 to 2.04~mJy in May 2003
and that its 8.4~GHz flux density increased from 2.21~mJy to 3.36
mJy between Feb and May 2003, as it appeared to increase
dramatically in size.

\section{Discussion}
\label{sec:dis}

\subsection{Does Arp~299 host one or more AGNs?}
\label{sec:agn}

  Any AGNs in Sources A, B1, C, C$'$ or D are buried in many 
magnitudes of extinction at visible and IR wavelengths \citep{aah}.
Sources A and B1 are thought to be the nuclei of the two galaxies which
are merging, and both are associated with hard X-ray point sources 
\citep{bal03,zez03}, so they seem to be likely AGN candidates.
[\citet{bal03} finds hard X-ray emission and \citet{zez03} 
show that it is localized at A and B1.]
VLBI observations on baselines of thousands of kilometers 
are able to detect even optically obscured AGNs,
because VLBI isolates only high brightness temperature 
($T_{b}>10^6$--$10^{7}$~K) objects. In contrast,
\hii\ regions must 
have $T_{B} < 10^{4}$, and supernova remnants generally have
low brightness temperatures as well.
However, high brightness temperature is not an exclusive discriminator
for AGNs, since some young radio SNe also are detected on milliarcsecond scales
(e.g., Bietenholz, Bartel, \& Rupen 2001; Perez-Torres et al. 2002; 
Bartel \& Bietenholz 2003),
implying brightness temperatures similar to some AGNs.

There have been a number of suggestions in the literature, based on
observations in a variety of frequency bands, 
that source A may harbor an AGN \citep{vlbi,lai,aah,zez03}.
In fact, the total VLA-scale radio power of Source A is considerably 
higher than the cores of most low-luminosity AGNs (LLAGNs) and classical 
Seyfert galaxies \citep{ho01,sch01,the01,nag02}.  Even if an AGN is
present, it is clear from the optical and infrared imaging that
Source A contains a strong star-formation complex.  The inferred
supernova rate in this complex, derived from radio flux densities
and the strength of near-infrared [Fe II] emission, is
0.6--0.7~yr$^{-1}$ \citep{aah}, while the VLA radio flux 
density implies a supernova rate as high as several per year
(see Section ~\ref{sec:vlacompt}).  Strong CO emission \citep{sar91,casoli}
and the lack of a strong optical nucleus \citep{aah} 
at the location of source A confirm that it is heavily 
enshrouded by dust.  

Previous VLBI measurements by \citet{vlbi} detected Source A with
a correlated flux density of 1.4~mJy at 
1.6~GHz on a transatlantic VLBI baseline, implying a source size 
$\leq$ 1~pc with a radio power of 3$\times 10^{20}$ W Hz$^{-1}$ and 
a brightness temperature $\geq$ 5$\times 10^{7}$~K.
\citep{smi98a} detected source A with correlated flux densities
of 15 and 7~mJy at 1.6~GHz, on respective projected 
baselines of about 50~km and 100~km, implying that there
is considerable structure on scales between $\sim$ 0\dsec005 
and $\sim$ 0\dsec 5.  Therefore, any AGN that might be present  
is embedded in a larger region of star formation and 
probably does not make a significant contribution to
the radio power of Source A.  Other star-formation
complexes on similar scales have been detected with VLBI 
observations: Arp~220 \citep{smi98b}, which contains an
AGN but not an energetically significant one, and Mrk~231 
\citep{tay99}, which contains a strong AGN.
 
Our detection of five compact radio sources with $T_{B}>10^7$K 
in Source A makes it unlikely that all the milliarcsecond-scale
radio emission is AGN-powered, since 
it is highly unlikely that there are 
five AGNs within Source A. However, there could be a single 
AGN present, as inferred from the hard X-rays \citep{bal03,zez03}.
In this regard, it is notable that source A1 is
the only milliarcsecond ($\lesssim$1~pc) compact object with an optically
thin synchrotron spectrum; if there is an AGN present, it mostly likely
is Source A1.  If Source A1 were an AGN, it would
be intriguing to have an apparent young supernova
(A0) only 2.6~pc away from the AGN.  

The X-ray point source associated with 
Source A has a 0.1--10~keV luminosity of $4\times 10^{39}$~ergs~s$^{-1}$
\citep{zez03}.
If the X-ray emission from Source A
were assumed to be associated with a single one of its milliarcsecond
radio sources, such as Source A1, we could estimate the radio to X-ray
ratio for that source.  For Source A1, 
$R_X = \nu P_\nu\ ({\rm 5\ GHz})/L({\rm 2-10\ keV})$ would
be $\sim 2\times 10^{-3}$, consistent
with values found for several other LLAGNs imaged on milliarcsecond
scales \citep{ulho01b} and also consistent with the values found for
the nucleui in the merger system NGC~3256 \citep{nef03}.
Source B1 has hard X-ray emission suggestive of an 
AGN \citep{bal03,zez03}, somewhat weaker than A, but no compact AGN is 
detected at radio wavelengths.  Still, the radio to X-ray ratio for the
milliarcsecond source in B1 is consistent with the range of values found
for LLAGNs.  
Sources C, C$'$, and D have much softer X-ray spectra \citep{zez03},
as well as no high brightness radio emission, suggesting that they
harbor no AGNs.  
Although the radio observations alone do not determine whether there 
are AGNs in Arp~299, we are able to say that any AGN present are unlikely
to play an important role in the merger's overall energy balance.

\subsection{Young Supernovae Within Source A - A Supernova Factory}
\label{sec:sourceA}

Source A is particularly strong; its VLA 8.4~GHz radio power
of 1.8$\times 10^{22}$ W Hz$^{-1}$ is more than 50 times the 
strongest non-thermal source 
in NGC~4038/9 \citep{neff00} and $\sim$27,500 
times the power of Cas A at the same wavelength.
Source a contains at least three distinct near-infrared sources, possibly
young star clusters, in its central 0.86 arcseconds \citep{aah}.
The present VLBI imaging reveals at least five compact radio emitters
within Source~A.  Four of these five have flat or inverted radio
spectra between 2.3 and 8.4~GHz, a property that is characteristic of
young Type II supernovae near their peaks \citep{wei02}.  Although such 
spectra also may be indicative of thermal gas associated with young star 
clusters, such clusters would have
brightness temperatures $\leq 10^4$~K, and would not be detected by
VLBI observations.  

Source A0 has an extremely inverted
spectrum (with $\alpha \geq +1.8$, Figure~\ref{fig:vlbaspec}).
Such an inverted spectrum is characteristic
of very young supernovae in the early days after their explosion, 
when the supernova blast has not yet completely penetrated the 
circumstellar mass-loss shell, which remains optically thick 
to lower frequency radio emission for months in ``normal" SN 
environments, and for years in very dense surroundings \citep{che82,
wei02}.  The spectra of Sources A2, A3, and A4
are typical of young SNe and Source A1 has a spectrum suggesting
an older SN remnant which has become optically thin down to 
frequencies of 1.4~GHz.

If we assume that all five VLBI sources within A are young supernovae,
they are spread over a total extent of 90~pc (440~mas). 
The three sources A0, A1, and
A2 are much more tightly clustered, with projected separations of only 
2.6~pc (13~mas) between A0 and A1, and 10.4~pc (52~mas) between A0 and A2.
Assuming that these sources all are young supernovae, it is quite
possible that A0 and A1, and perhaps even A2, are found within the
same cluster of young stars.  Super star clusters seen optically in
other merging galaxies have canonical radii of $\sim 4$~pc, but
may be as large as tens of parsecs (e.g., Whitmore et al. 1999).
\citet{aah} infer that Source A had the peak of its
starburst 6--8~Myr ago. This corresponds to the main-sequence lifetime
of O stars of $\sim$25--30$M_\odot$ \citep{sch92,sch93}.  Massive stars
formed in that burst should just now be ending their lives
on the main sequence, and the radio sources that we see now probably 
result from the interactions of these explosions with their surroundings.
{\it We should therefore expect to see frequent  
supernova explosions from this ``supernova factory.''} 

 The five identified 
VLBI sources account for only 6\% of the total VLA 8.4~GHz flux density of
87~mJy in Source A, which has an optically thin synchrotron
spectrum on a scale of hundreds of milliarcseconds (cf. 
Table~\ref{tab:vlafits}).  It is likely that most of the
radio emission not detected on milliarcsecond scales 
results from earlier supernovae that have now expanded
and faded
beyond the detection range of VLBI and merged into the
smoother interstellar medium.  The expansion to size scales 
larger than $\sim 0.2$~pc typically takes a hundred years or so
in a ``normal''  interstellar medium, while the general merging 
into that interstellar medium may take $\sim 10^5$~yr \citep{condon92}.  
However, these time scales are quite uncertain in the denser 
environment of Arp~299.

If Source A does contain five young supernovae, why are none
detected in sources B1, C, or C$'$?   \citet{aah} use infrared
fluxes to infer supernova rates 0.6--0.7~yr$^{-1}$
for Source A, but find supernova rates for B1, C, and C$'$  
that are factors of 5--20 below that for Source A. 
We derive similar rates based on the VLA-scale radio emission
(see Section ~\ref{sec:vlacompt}).
It seems reasonable that the number of detectable radio 
supernovae in B, C, and C$'$ should therefore be no more 
than 5\%--20\% of the number in Source A, or fewer than one 
per region for our most sensitive VLBI observations. 
If the supernovae in A tend to be stronger radio
emitters due to a denser interstellar environment, as we expect, 
the number of supernovae detectable in the other sources 
would be even smaller.  
Sources B1,
C, and C$'$ are also thought to be younger starbursts \citep{aah},
so they may be just beginning significant supernova activity.

Another alternative for the compact radio sources in source A
is that they are X-ray binaries or microquasars.  However, X-ray
binaries typically have radio to X-ray ratios $R_X\lesssim 10^{-5}$ 
\citep{fen01}.  Therefore, we would expect 
$\gtrsim 10^{42}$~ergs~s$^{-1}$ in X-rays from Source~A if
the compact radio sources were X-ray binaries, 
compared to a measured value of 
$\sim 1.3\times 10^{40}$~ergs~s$^{-1}$ \citep{zez03}.
This appears to rule out active binaries as the origin of the 
milliarcsecond radio sources.

Future long-term VLBI monitoring of Arp~299 should enable us to 
test the idea that Source A contains a cluster of young
SNe.  First, new supernova explosions
could be discovered as they occur, providing a direct measure of
the radio supernova rate and indicating 
whether or not {\it all} supernovae in Source A are likely to 
be radio loud.  If the detected radio supernova rate were one
per decade rather than one per year or so, much less than the
total estimate of $\sim 0.6$~yr$^{-1}$ \citep{aah}, then this
result would indicate strongly that only a fraction of the 
Arp~299 supernovae are radio loud, at least above the VLBA
detection threshold.
Second, the fading of the young supernovae could
be measured, for comparison with
observations of young supernovae in less dense environments;
this would provide useful constraints on 
the properties of the star-forming regions,
as well as enabling an estimate of the age of the young radio
supernovae already detected in Source A.
Third, we should be able to verify the nature of
Source A1:   it should fade monotonically 
or rise and fall coherently if it is a supernova, while more
stochastic flux variations would suggest that it is an AGN.

\subsection{The Supernova Rate for VLA Sources A, B1, C and C$'$ }
\label{sec:vlacompt}

Radio sources with ``normal'' nonthermal spectral slopes 
(i.e., $\alpha < 0.0$) are likely to be complexes of 
older supernova remnants, with the radio emission 
dominated by optically thin synchrotron emission.  
Inspection of Tables~\ref{tab:matched} and \ref{tab:vlaprop}, 
where the resolutions at 
all wavelengths have been matched to the 1.4~GHz beam, shows that 
sources A, B1, C, and C$'$ all have such spectra on arcsecond
scales.  The spectral indices of $-0.5$ to $-0.65$ are consistent
with typical galactic supernova remnants \citep{wei02}, allowing
us to use the observed flux and the spectral index
to derive the supernova rate in each of these 
arcsecond-scale ($\sim$100$-$200 pc) sources.

There has been considerable discussion about the supernova rate 
corresponding to a given radio flux density, with estimates 
varying by a factor of 10 or so.  
In studying NGC~4038/9, \citet{neff00} pointed out that very different 
supernova rates will be derived, depending on whether
the global radio emission of the merger, or only the emission 
identifiable with compact radio sources, was considered.  This 
is because the bulk of the total radio emission of a starburst 
galaxy is caused by synchrotron-emitting electrons that have 
escaped the vicinity of the supernova remnants that
accelerated them, and diffused throughout the galaxy.
In NGC~4038 \citet{neff00} obtained the same supernova rates using 
the approach in \citet{jsu82} and using the technique described by 
\citet{condon92}, as long as the latter approach was applied to 
the radio emission from the {\it entire} galaxy. 
In Arp~299, as in NGC~4038/9 \citep{neff00}, we consider 
only radio emission that is compact (arcsecond scale or smaller), 
and therefore derive the supernova rate for the VLA-scale
sources using Equation (7) from \citet{jsu82} 
(equation~\ref{eqno2}, below). 

For Sources A, B1, C and C$'$, we derive the supernova rate by 
assuming supernova blast wave energies of $10^{50}$~ergs and 
ambient densities of 1~cm$^{-3}$.  
In this case, the supernova rate 
$R_{\rm SN}$ is given by:

\begin{equation}
R_{\rm SN} =0.37 f \times \biggl(\frac{S_{\rm 4.9\ GHz}}
{5.0\ \mathrm{  mJy}}\biggr)\biggl(\frac{408}{4885}\biggr)^{\alpha+0.75}\ {\rm 
yr}^{-1}\ .
\label{eqno2}
\end{equation}
Here, $f$ is the fraction of the radio flux density that can still
be associated with individual supernova remnants on parsec scales.
For the 4.9~GHz flux densities, we use the value measured 
in the matched angular resolution images (Table~\ref{tab:matched}).
For the spectral index $\alpha$ between 408~MHz and 4.9~GHz, we
use the measured value between 4.9~GHz and 8.4~GHz (Table~\ref{tab:vlaprop}),
thus assuming that an ensemble of supernova remnants unresolved by
the VLA has a net spectrum that is straight between 408~MHz and 8.4~GHz.
The resulting estimates of the supernova rates required to
produce the observed radio flux densities for  
sources A, B, C, and C$'$, are given in Table~\ref{tab:vlaprop}.

The total radio emission from source A, which corresponds to a
power $2.7\times 10^4$ times that of Cas A,
requires a supernova rate of $\sim 4.7f$ yr$^{-1}$.  The nonthermal
sources in B1, C, and C$'$ are considerably less powerful, with their
supernova rates estimated at $0.3f$--$1.0f$~yr$^{-1}$.  As stated in
Section 5.2, the five identified VLBI sources account for only 6\% of
the total flux density at 8.4~GHz, so a lower limit for $f$ is 
$\sim 0.06$ for Source A.  Since there may be other emission
on scales of several parsecs that is too big to be seen in the
milliarcsecond VLBI imaging, a reasonable value for $f$ may be
0.1--0.2, which then would provide a calibration similar to those
advocated by \citet{condon92} and \citet{hua94}.  This would 
imply a supernova rate of 0.5-1.0~yr$^{-1}$ for Source A, in
good agreement with the result found by \citet{aah}.

\subsection{What is Source D?}
\label{sec:sourceD}

\begin{figure}
\figurenum{1}
{\plotone{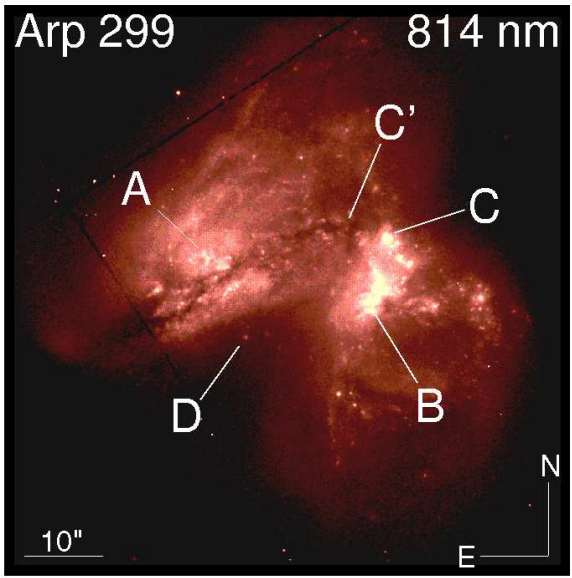}}
\caption{{\it HST} WFPC2 814 nm image, made from public archival data. 
North is up, East is left.  The two components of the merger are clearly 
distinguishable.  Radio source A is located in the center of the 
eastern galaxy; its IR luminosity is $\sim$50\% of the 
total IR luminosity of the galaxy pair.
Sources B and C are in the central region
of the western galaxy, B lower and C upper.
This image 
shows approximately the same field of view as the radio 
images in Figures 2-6.  }
\label{fig:vis}
\end{figure}

\begin{figure}
\figurenum{2}
{\plotone{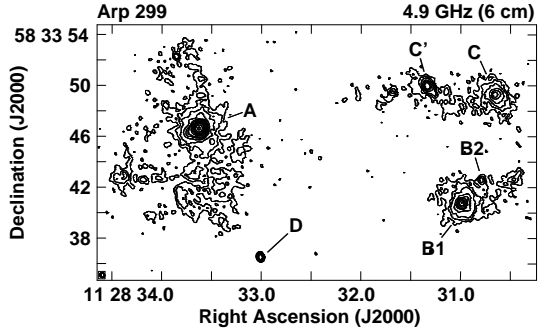}}
\caption{4.9~GHz (6~cm) VLA image of Arp~299.  
Radio sources A, B, C, C$'$, and D are indicated. 
Contour levels are $-$0.065, 0.065, 0.120, 0.2, 0.4, 
0.7, 1, 2, 4, 7, 10, 20, and 40~mJy~beam$^{-1}$,  
the peak flux density is   46.0~mJy~beam$^{-1}$, 
the rms noise is       15.7 $\mu$Jy~beam$^{-1}$, and
the restoring beam size is 
0\dsec 38 $\times$ 0\dsec 31 with position angle of $17^\circ$.}
\label{fig:6cm}
\end{figure}

\begin{figure}
\figurenum{3}
{\plotone{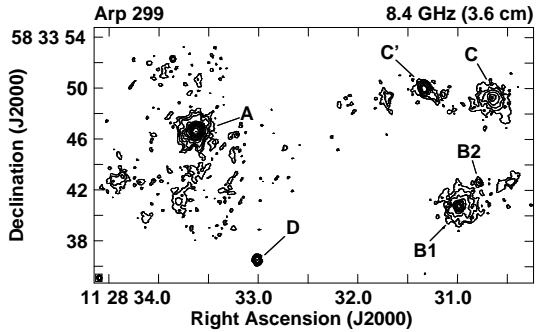}}
\caption{8.4~GHz (3.6~cm) VLA image of Arp~299.  Contour
levels are $-$0.070, 0.070, 0.121, 0.200, 0.400,
0.700, 1, 2, 4, 7, 10, 20, and 39 mJy beam$^{-1}$,
the peak flux density is   39.1 mJy beam$^{-1}$,
and the rms noise is   14.1 $\mu$Jy beam$^{-1}$. 
The data have been weighted to match the 4.9~GHz image shown
in Figure 2, with a restoring beam of
0\dsec 38 $\times$ 0\dsec 31 with position angle of $17^\circ$.}
\label{fig:4cm}
\end{figure}

\begin{figure}
\figurenum{4}
{\plotone{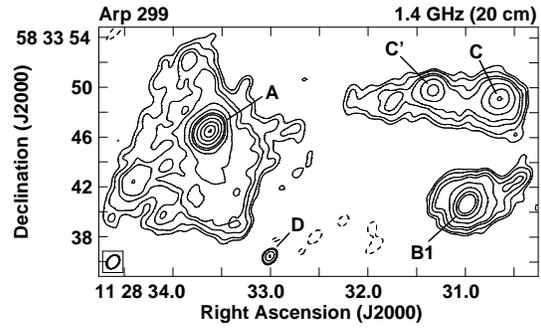}}
\caption{1.4~GHz (20~cm) VLA image of Arp~299.
Contour levels are $-$0.27, 0.27, 0.4, 
0.7, 1, 2, 4, 7, 10, 20, 40, 70, and 100 mJy beam$^{-1}$, 
the peak flux density is 101.2~mJy~beam$^{-1}$, the rms noise is 69.2 
$\mu$Jy~beam$^{-1}$, and the restoring beam size is 
1\dsec 21 $\times$0\dsec 88 in position angle of $-38^\circ$.
Extensive diffuse emission is clearly present, even in these A 
configuration observations.}
\label{fig:20cm}
\end{figure}

\begin{figure}
\figurenum{5}
{\plotone{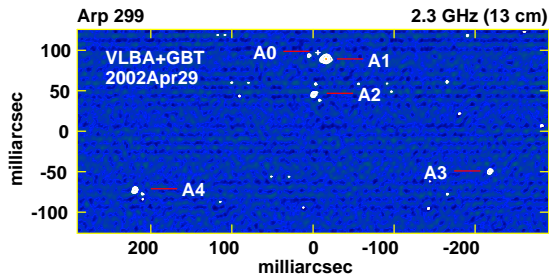}}
\caption{2.3~GHz image of Source A, using the VLBA together 
with the GBT, from data acquired on 2002 April 29.
Contour levels are at ($-$3, 3, 6, 24) times the rms noise of 
$3.7\times 10^{-5}$ mJy~beam$^{-1}$, and the peak flux density is 
1.76~mJy~beam$^{-1}$.  The coordinate of (0,0) mas corresponds to
$(\alpha,\delta)$ = (11$^h$28$^m$33.62201$^s$, 58$^\circ$33\arcmin 46\farcs 
6100).  VLBI sources A1, A2, A3, and A4 are indicated.
The cross marks the position of source A0, not detected at 2.3~GHz.  The beam
size is $5.58\times 4.49$~mas in position angle $-28^\circ$.}
\label{fig:vlba13cm-wide}
\end{figure}

\begin{figure}
\figurenum{7}
{\plottwo{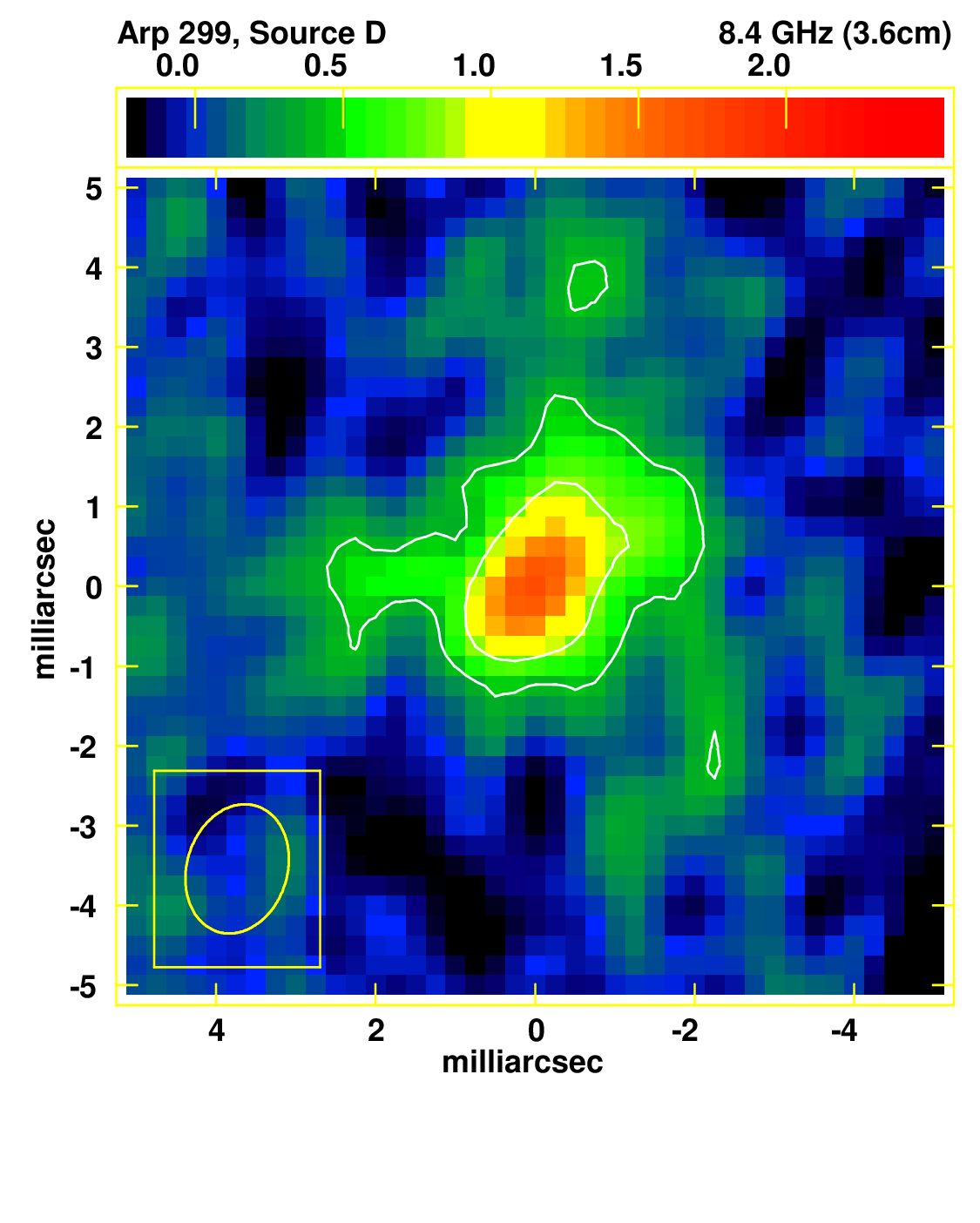}{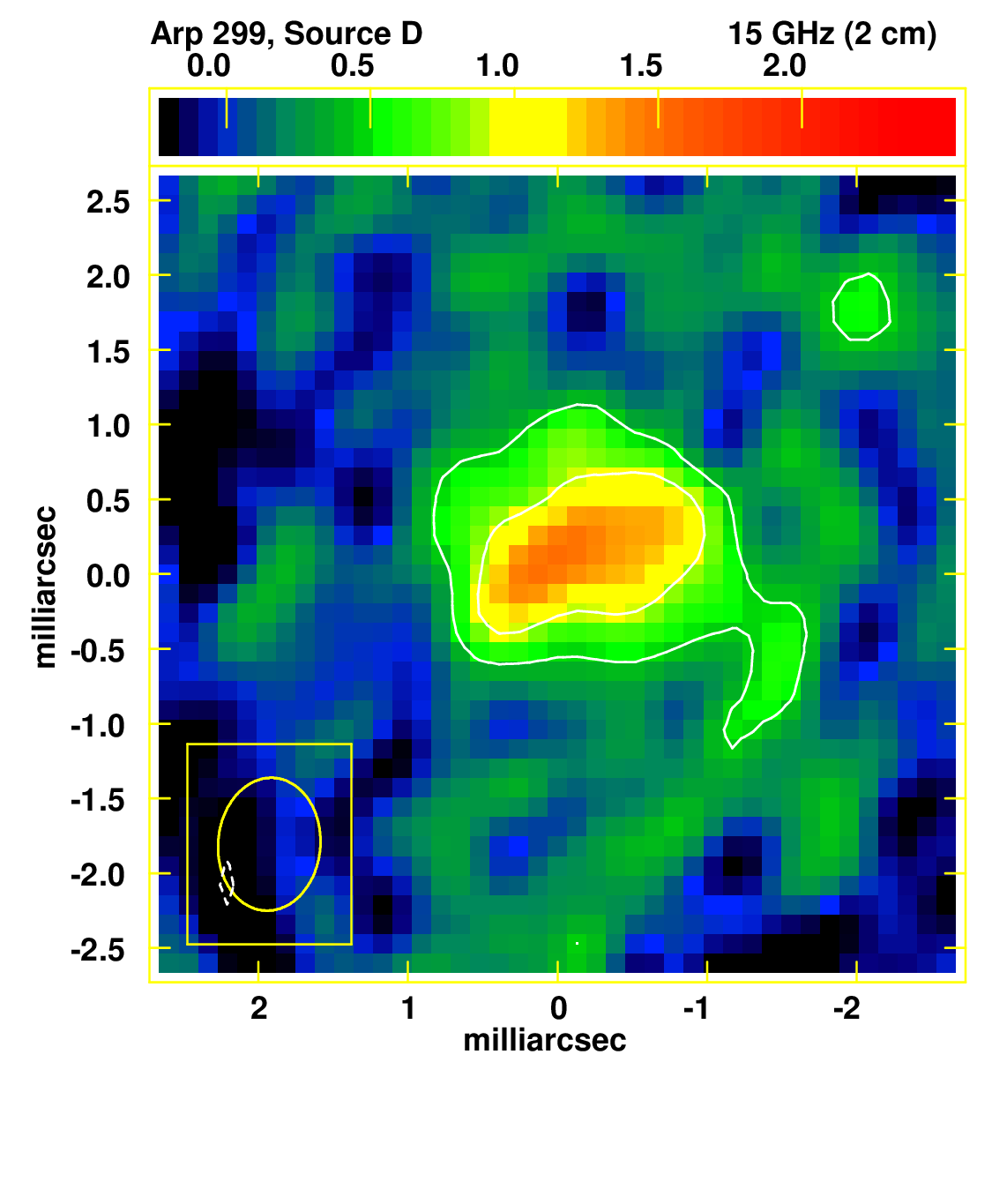}}
\caption{VLBA images of source D from 2003 May 1.  The previously
(February 2003) unresolved Source D is clearly extended in these 
observations, with a size of $0.35 \times 0.22$ pc.  The left panel 
is an 8.4~GHz (3.6~cm) image, with a peak flux density of 1.71~mJy~beam$^{-1}$;
contour levels are at ($-$3, 3, 6) times the rms noise of 
0.137 mJy~beam$^{-1}$.  The right panel is a 15~GHz (2~cm) image
with a peak flux density of 1.53~mJy~beam$^{-1}$;
contour levels are at ($-$3, 3, 6) times the rms noise of
0.152 mJy~beam$^{-1}$.}
\label{fig:vlba-d}
\end{figure}

Radio source D does not appear to be in the main interaction region of 
Arp~299 (see Fig.~\ref{fig:vis} or \ref{fig:6cm}). Therefore, it 
is worth considering the
possibility that Source D actually is not associated with the galaxy, 
but is a background source instead, as suggested previously \citep{zhao}.
Using the VLA in 1980, \citet{condon} 
reported a continuum radio source in the vicinity of source D at 
1.4~GHz, with a flux density of $\sim$4~mJy in a 2$\arcsec$ beam.  
This flux density is much higher than the 1~mJy upper limit in 1989
\citep{hua90} the 0.8~mJy flux density that we derive from 1991 
archival data (Figure 5, Tables 3 and 4) the 0.5~mJy measured in
a snapshot observation on 2003 April 29, or the VLBI upper limit
from this work.  Source D might be variable at 1.4~GHz, 
but we consider it more likely that the measurement by \citet{condon}
may have included more diffuse emission surrounding the source - 
we note that the VLA was in an early, hybrid, array at the time
of these observations.

It is possible to use the many individual observations to 
investigate the
possible variability of Source D; comparisons to the other
compact sources can be used to infer the reality of any
apparent variability.   Therefore, we have measured the flux
densities of sources A, B1, C, C$'$ and D in independent data 
sets observed between 1989 and 2003; results are 
shown in Table~\ref{tab:var} and in Fig.~\ref{fig:vlavar} 
{\footnote{ \citet{hua90} reported a flux density of 1.8~mJy 
from the source at 8.44~GHz on 1990 March 1; this actually 
is a composite of two data sets a few days before and after 
that date, which we have reprocessed  individually.  \citet{zhao} 
detected the source at 8.2~GHz in March 1993 (B array) with a flux 
density of 2.2$\pm$0.1~mJy; this data is in good agreement with
nearby data we did use. }}. 
 We conservatively use only data 
from the high-resolution A and B configurations
at 4.9 and 8.4~GHz, where we expect diffuse radio emission 
to be resolved out; this approach insures that 
measurements of the relatively isolated source D are not 
contaminated by more diffuse emission.  Images were made with 
matched restoring beams for each array/frequency combination used,
to assure that measurements were made with the same resolution.
Sources A, B1, C, and C$'$ were all found to be extended, and were
fitted with single Gaussian sources which were consistent in size 
and orientation in all observations made with a particular array.
Source D was found to be unresolved in all individual observations, 
and was measured as a point source.  Flux density errors were determined 
as discussed in section 4.1, and are approximately the size of the
points shown in Figure \ref{fig:vlavar}.

In a fixed configuration, sources A and B1 do not vary significantly 
at 4.9~GHz and 8.4~GHz during 
the 13 years of observations.  
Sources C and C$'$ appear to vary, but 
they are very well resolved; we attribute their apparent variation to
differing resolutions and aperture plane coverages in (mostly) short
observations.  Source D clearly has changed in flux density at
4.9 and 8.4~GHz (Fig. \ref{fig:vlavar});  
the measured flux densities in 1990 and 1991 are consistent
with one another, but the more recent measurements show significantly
higher flux densities.  This trend is completey different from Sources
A and B1, leading us to infer that the variability of Source D is real
and is not a data-reduction artifact.

On VLBI scales,  Source~D underwent a dramatic flux-density 
increase at 2.3~GHz between 2002 and 2003, as well as becoming 
significantly resolved and increasing in 
flux at 8.4~GHz  in the three-month interval between 2003
February 9 and 2003 May 1 (see Figure~\ref{fig:vlba-d}).  One can imagine 
several possible explanations for this change.  If Source D is located 
within Arp~299, it may well have had a new supernova explosion in
early 2003, or a supernova explosion that suddenly has broken
through the circumstellar shell to become visible at the lowest VLBI
frequency.  If it is a background quasar, it may have ejected 
a new ``jet'' component in a radio flare.  However, one might
expect such a flare to give rise to a much stronger variation
at the higher frequencies, rather than increasing in flux by 
more than 600\% at 2.3~GHz.
Further, the apparent source size has increased by $\sim 1$~mas
in three months.  Even at a modest assumed redshift of 
$z\sim 0.1$, this would correspond to a large (but not unprecedented)
expansion speed of $v_{\rm app}/c\sim 26$.  

Our tentative conclusion is that Source~D is
located within Arp~299, probably within a very compact young 
star cluster.  Supernova explosions in Source~D would then account for
the apparent flux-density variability seen in the multi-epoch
VLA data, as well as for the rapid ``expansion'' and 2.3-GHz 
flux density increase in 2003.  In support of this conclusiton,
we note that in 2001 July, Source
D had a soft X-ray spectrum consistent with being a radio supernova
\citep{zez03}.  This conclusion can be validated 
by future VLBI imaging of Source~D.  If the source is a
relativistic jet, its size should continue to expand, whereas
a newly emerging supernova will stay in the same apparent
location, yielding a relatively constant size as a function
of time.

\section{Summary}
\label{sec:concl}

In this paper, we have discussed the five brightest radio 
sources in Arp~299.  Four of the five brightest sources, A, 
B1, C, and C$'$, have VLA spectra characteristic of complexes of 
supernova remnants, with inferred supernova rates ranging from a 
few tenths to a few supernovae per year.   VLBI
imaging reveals five compact radio sources within A, most of
which have flat or inverted radio spectra.   Based on the sizes 
and spectra, we identify these as probable young supernovae or 
supernova remnants in a ``supernova factory,'' with individual 
powers of hundreds to a thousand times that of Cas~A.  
One compact source (A0) has an extremely inverted radio spectrum,
with $\alpha\approx +2$ or greater between 2.3 and 8.4~GHz. Three 
of the compact VLBI sources are located within a region with 
a projected size of 10~pc and may well be within the same 
super star cluster or cluster complex.  An interesting alternative
possibility is that the one steep-spectrum source (A1) is a 
low-luminosity AGN, occurring at a separation of
only 2.6~pc (projected) from a very young supernova (A0).
The VLBI sources include less than 10\% of the total
VLA flux density, implying that much of the radio power in 
Source A comes from older supernova remnants that have 
expanded and merged into the surrounding ISM.  We do not
know which (if any) of the separate VLBI sources in Source A produce the 
observed X-ray emission. If we attribute the bulk of the
X-ray emission to one or more of the VLBI sources within Source~A,
the radio/X-ray ratio is far too high for all the compact objects to 
be beamed or unbeamed ``normal'' X-ray binaries or microquasars. Instead,
the total radio/X-ray ratio is consistent with a mix of young supernovae,
supernova remnants, and perhaps an AGN.

Radio source D, located at a projected distance of 2~kpc from source
A, is quite compact on VLBI scales, with the VLBI source apparently
containing all of the flux detected in the highest resolution VLA images.  
Its VLA 8.4~GHz and 4.9~GHz flux densities have both varied 
by factors of $\sim 2$ over the last 13 years.  On VLBI scales, the
Source D flux density at 2.3~GHz increased by a factor of $\sim 6$
between 2002 April and 2003 May, and its 8.4~GHz flux density increased
by a factor of 1.6 in three months in early 2003; during
early 2003, Source D also increased in size by at least a factor
of 2 at 8.4~GHz.  Although Source D could be a background quasar
undergoing superluminal expansion, the observed properties fit more
closely to a star-formation complex.  VLBI monitoring should allow
us to verify the SN hypothesis (or not) for Source D, and a high-resolution 
mid-infrared image of Source D would help confirm its identity 
as a young cluster in Arp~299.

\acknowledgments

We thank Anneila Sargent and Dan Weedman for helpful discussions.
Stacy Teng thanks the NSF for supporting some of this work
through their Research Experience for Undergraduates program
at NRAO, and Susan Neff acknowledges partial support via NASA 
grant 344-01-71-21.  This research has made use of
NASA's Astrophysics Data System Abstract Service and the NASA/IPAC
Extragalactic Database (NED) which is operated by the Jet Propulsion
Laboratory, California Institute of Technology, under contract with 
the National Aeronautics and Space Administration.  

%
%



%
%

\begin{deluxetable}{ccccccc}
\tabletypesize{\scriptsize}
\tablecolumns{7}
\tablewidth{0pc}
\tablecaption{VLA Observations of Arp~299 Used in Combined Images}
\tablehead{
\colhead{Observing}&\colhead{Frequency}&\colhead{Array}&
\colhead{Integration}&\colhead{Phase}&
\colhead{Derived Flux of}&\colhead{Image rms}\\
\colhead{Date}&\colhead{}&\colhead{Configuration}&
\colhead{Time}&\colhead{Calibrator}&\colhead{Phase Calibrator}
&\colhead{at Robust=0}\tablenotemark{a}\\
\colhead{}&\colhead{(GHz)}&\colhead{}&\colhead{(min)}&
\colhead{(J2000)}&\colhead{(Jy)}&\colhead{($\mu$Jy beam$^{-1}$)}\\}
\startdata
\cutinhead{1.4 GHz (20 cm) observations}
1991 Aug 3 & 1.46 & A & 143 & J1219+4829 & 0.57& 57\\
\cutinhead{4.9 GHz (6 cm) observations}
2000 Oct 24 & 4.86 & A & 177 &J1128+5925& 0.46& 20.1\\
1994 May 16 & 4.86 & BnA & 5&J1146+5356 &0.37& 97.9\\ 
1993 Dec 19 & 4.86 & D & 14 &J1219+4829&0.62& 50.1\\ 
1993 Jun 11 & 4.86 & CnB & 19 &J1219+4829&0.60& 61.7\\ 
1993 May 7 & 4.86 & B & 15 &J1219+4829&0.62& 54.3\\ 
1993 Mar 13 & 4.86 & B & 25 &J1219+4829&0.62\tablenotemark{b}& 46.4 \\  
1993 Jan 28 & 4.86 & BnA  & 26 &J1219+4829&0.61& 75.3\\ 
1992 Aug 28 & 4.86 & D & 145 &J1146+5848&0.37& 28.3\\ 
1992 Apr 3 & 4.86 & C & 170 &J1146+5848&0.37& 23.4\\ 
1991 Jul 5 & 4.86 & A & 35 &J1219+4829&0.49& 38.2\\ 
1990 May 9 & 4.86 & A & 25 &J1219+4829&0.62& 55.6\\ 
1989 Jan 14 & 4.86 & A & 7 &J1146+5848&0.44& 73.2\\ 
\cutinhead{8.4 GHz (3.6 cm) observations}
2001 Mar 19 & 8.46 & B & 171&J1128+5925&0.28& 17.3\\
1999 Apr 7 & 8.46 & D & 31&J1219+4829&0.61& 44.3\\ 
1999 Jan 21 & 8.46 & C & 32 &J1219+4829&0.68& 33.7\\ 
1994 May 16 & 8.44 & BnA & 15&J1146+5356&0.34& 29.7\\ 
1994 Feb 18 & 8.44 & D $\rightarrow$ A & 51&J1219+4829&0.66& 26.4\\ 
1993 Dec 19 & 8.44 & D & 13 &J1219+4829&0.65& 29.3\\
1993 Jun 11 & 8.41 & CnB  &19&J1219+4829&0.63&46.7\\
1993 May 7 & 8.44 & B & 15 &J1219+4829&0.64& 31.2\\
1993 Mar 13 & 8.41 & B& 25&J1219+4829&0.66&37.1\\
1993 Jan 28 & 8.44 & BnA & 21 &J1219+4829& 0.63& 25.5\\ 
1991 Jul 5 & 8.44 & A & 32 &J1219+4829&0.51& 34.2\\ 
1991 Jun 22 & 8.42 & A & 50 &J0958+6533&1.13& 37.5\\
1990 May 9 & 8.44 &A &42&J1219+4829&0.62&37.1\\ 
1990 Mar 4 & 8.44 & A & 15 &J1217+5835& 0.45& 50.1\\ 
1990 Feb 24 & 8.44 & A & 15 &J1217+5835&0.47& 30.5\\
\enddata
\tablenotetext{a}{Robust = 0 indicates that data weighting intermediate
between pure natural weighting (highest sensitivity) and pure 
uniform weighting (highest resolution)  was
used in producing the image  \citep{bri95,brig98}. }
\tablenotetext{b}{The flux of this phase calibrator 
was estimated from those derived from the other archived 
data sets taken near the time of this observation.  The original 
observation lacked an amplitude calibrator.}
\label{tab:vlaobs}
\end{deluxetable}

\begin{deluxetable}{ccccccc}
\tabletypesize{\scriptsize}
\tablecolumns{7}
\tablewidth{0pc}
\tablecaption{VLBI Observations of Arp~299}
\tablehead{
\colhead{Observing}&\colhead{Frequency}&\colhead{Array}&\colhead{Data}&
\colhead{Integration}&\colhead{Image rms}&\colhead{Beam Size}\\
\colhead{Date}&\colhead{}&\colhead{Configuration}&\colhead{Rate}&
\colhead{Time}&\colhead{at Robust=5\tablenotemark{a}}&
\colhead{at Robust=5\tablenotemark{a}}\\
\colhead{}&\colhead{(GHz)}&\colhead{}&\colhead{(Mbit s$^{-1}$)}&
\colhead{(min)}&\colhead{($\mu$Jy beam$^{-1}$)}&\colhead{(mas)}\\}
\startdata
2002 Apr 29 & 2.27 & VLBA+GBT & 256 & 330 & 35 &$5.58\times 4.49$\\
2003 Feb 09 & 8.41 & VLBA & 256 & 438 & 51 &$1.84\times 1.27$\\
2003 May 01 & 2.27 & VLBA& 128 & 58 & 162 & $6.19\times 4.64$ \\
2003 May 01 & 4.98 & VLBA & 128 & 53 &129 & $2.56\times 1.87$ \\
2003 May 01& 8.41 & VLBA& 128 & 53 & 137 & $1.66\times 1.26$ \\
2003 May 01 & 15.36 & VLBA& 128 & 167 & 152 & $0.83\times 0.67$ \\
\enddata
\tablenotetext{a}{Robust = 5 indicates that pure natural weighting 
of the data was used in producing the images \citep{bri95,brig98}. 
}

\label{tab:vlbaobs}
\end{deluxetable}

\begin{deluxetable}{ccccccccc}
\tabletypesize{\scriptsize}
\tablecolumns{9}
\tablewidth{0pc}
\tablecaption{High Resolution Measurements of the Five Compact VLA Sources}
\tablehead{\colhead{Source} & \colhead{RA} & \colhead{Dec} & 
\colhead{$S\tablenotemark{a}({\rm 1.4\ GHz})$} 
& \colhead{$S\tablenotemark{a}({\rm 4.9\ GHz})$}  & 
\colhead{$S\tablenotemark{a}({\rm 8.4\ GHz})$}
& \colhead{Major Axis\tablenotemark{b}} & \colhead{Minor 
Axis\tablenotemark{b}} &\colhead{P.A.\tablenotemark{b}}\\
\colhead{} &\colhead{(J2000)} &\colhead{(J2000)} &\colhead{(mJy)} 
&\colhead{(mJy)} 
&\colhead{(mJy)} 
&\colhead{(arcsec)} 
&\colhead{(arcsec)} &\colhead{(deg)}\\}
\startdata
A&11 28 33.63&58 33 
46.67&163.3$\pm$8.2&101.1$\pm$5.0&77.2$\pm$3.9&0.38&0.30& 127 \\
B1&11 28 30.99&58 33 
40.78&41.6$\pm$2.1&11.1$\pm$0.6&9.1$\pm$0.5&0.27 &0.25& 100 \\ 
C&11 28 30.65&58 33 
49.29&21.8$\pm$1.1&6.9$\pm$0.4&5.9$\pm$0.3&0.83 & 0.74&32 \\ 
C$'$&11 28 31.33&58 33 
50.02&12.2$\pm$0.3&4.7$\pm$0.3&4.1$\pm$0.2&0.49 &0.36& 28 
\\ 
D\tablenotemark{b}
&11 28 33.01&58 33 
36.55&0.91$\pm$0.08&1.94$\pm$0.10&2.43$\pm$0.12&0.17& 0.11& 168 \\ 
\cutinhead{Image Properties}
 & & &  1.4~GHz & 4.9~GHz &  8.4~GHz &  & & \\
Restoring Beam & & & 1\dsec21$\times$0\dsec88 & 0\dsec38$\times$0\dsec31 & 0
\dsec38$\times$0\dsec31 & 
& & \\
Beam P. A.   & & & -38\deg     & 17\deg  &  17\deg  &   &   &  \\
rms ($\mu$Jy beam$^{-1}$)  & & & 67       &  17 &  13  &  &  &  \\
\enddata
\tablenotetext{a}{Flux densities derived from Gaussian fits to the 
radio components }
\tablenotetext{b}{Measured at 8.4~GHz }
\tablenotetext{c}{Source D is unresolved except possibly at 8.4~GHz.}
\label{tab:vlafits}
\end{deluxetable}

\begin{deluxetable}{cccc}
\tabletypesize{\scriptsize}
\tablecolumns{6}
\tablewidth{0pc}
\tablecaption{Matched Beamsize Flux Densities of the Five VLA Sources }
\tablehead{\colhead{Source} &\colhead{$S({\rm 1.4\ GHz})$} 
&\colhead{$S({\rm 4.9\ GHz})$} &\colhead{$S({\rm 8.4\ GHz})$} \\
\colhead{} &\colhead{(mJy)} &\colhead{(mJy)} &\colhead{(mJy)} \\} 
\startdata
A&163.3$\pm$8.2&115.0$\pm$5.8&87.1$\pm$4.4\\
B1&41.6$\pm$2.1&20.1$\pm$1.0&14.5$\pm$0.7\\
C&21.8$\pm$1.1&11.0$\pm$0.6&7.7$\pm$0.4\\
C$'$&12.2$\pm$0.6&6.7$\pm$0.3&5.1$\pm$0.3\\
D&0.91$\pm$0.08&1.73$\pm$0.09&2.45$\pm$0.32\\
\enddata
\label{tab:matched}
\tablecomments{All measurements made from images with 1.21 $\times$ 0.88 beam,
Position Angle -38 degrees E of N.}
\end{deluxetable}

\begin{deluxetable}{cccccc}
\tabletypesize{\scriptsize}
\tablecolumns{6}
\tablewidth{0pc}
\tablecaption{Derived Physical Properties of the VLA Sources A, B1, C, C$'$, D}
\tablehead{\colhead{Source}&\colhead{{${\alpha}_{5-8}$}}
&\colhead{$T_{B}$(8.4 GHz)}&\colhead{$P$(8.4 GHz)}
&\colhead{$P/P_{\rm CasA}$}&\colhead{SN Rate}\\
\colhead{}&\colhead{}&\colhead{($10^3$ K)}&\colhead{(W Hz$^{-1}$)}
&\colhead{}&\colhead{(yr$^{-1}$)}\\}
\startdata
A   &$-0.51\pm 0.13$ &17.1  &$1.8\times 10^{22}$ & 27,500 &0.5--1.0  \\
B1  &$-0.60\pm 0.13$ &3.4   &$2.9\times 10^{21}$  & 4580 &0.1--0.2  \\ 
C   &$-0.65\pm 0.13$ &0.2   &$1.6\times 10^{21}$ & 2430 &0.05--0.1  \\ 
C$'$&$-0.50\pm 0.13$ &0.6   &$1.0\times 10^{21}$ & 1610 &0.03--0.06  \\ 
D   &$+0.64\pm 0.13$ &3.3   &$4.9\times 10^{20}$ & 770 &\nodata  \\
\enddata
\label{tab:vlaprop}
\tablecomments{Spectral indices, total radio powers, and
supernova rates are measured from the matched-resolution data given in
Table~\ref{tab:matched}.  Peak brightness temperatures
are measured from the highest resolution data, given 
in Table~\ref{tab:vlafits}.  Supernova rates are derived using 
Equation (2) in Section 5.3, witn $f$=0.1--0.2.  No supernova rate
is estimated for Source D because its spectrum does not fit
the assumptions made in deriving Equation 2.  }
\end{deluxetable}

\begin{deluxetable}{cccccccc}
\tabletypesize{\scriptsize}
\tablecolumns{8}
\tablewidth{0pc}
\tablecaption{VLBI Component Flux Densities}
\tablehead{
\colhead{Source}&\colhead{RA}&\colhead{Dec}&
\colhead{$S({\rm 2.3\ GHz})$\tablenotemark{a}}&
\colhead{$S({\rm 8.4\ GHz})$\tablenotemark{a}}
&\colhead{Major Axis}&\colhead{Minor Axis}&\colhead{P.A.}\\
\colhead{Designation}&\colhead{(J2000)}&\colhead{(J2000)}&
\colhead{(mJy)}&\colhead{(mJy)}&\colhead{(mas)}&\colhead{(mas)}&
\colhead{(deg)}\\}
\startdata
A0 &11 28 33.6212&58 33 46.707&$<0.11$&$3.23\pm 0.28$& 
$<0.92$&$<0.64$&\nodata\\
A1&11 28 33.6199&58 33 46.699&$1.91\pm 0.19$&$0.60\pm 
0.09$&$<0.92$&$<0.64$&\nodata\\ 
A2&11 28 33.6218&58 33 46.655&$0.40\pm 0.05$&$0.67\pm 
0.09$&$<0.92$&$<0.64$&\nodata\\ 
A3&11 28 33.5931&58 33 46.560&$0.30\pm 0.05$&$0.39\pm 
0.08$&$<0.92$&$<0.64$&\nodata\\ 
A4&11 28 33.6502&58 33 46.538&$0.35\pm 0.05$&$0.40\pm 
0.08$&$<0.92$&$<0.64$&\nodata\\
D (2002 Apr 29)&11 28 33.0108&58 33 36.549&$0.30\pm 0.03$
& \nodata&$<2.79$&$<2.25$&\nodata\\
D (2003 Feb 9)&11 28 33.0108&58 33 36.549& \nodata
&$2.21\pm 0.20$&$<0.92$&$<0.64$&\nodata\\
D (2003 May 1)&11 28 33.0107&58 33 36.549&$2.04\pm 0.26$
&$3.36\pm 0.36$&1.8&1.1&141\\
\enddata
\label{tab:vlbafits}
\tablenotetext{a}{Except as noted, 2.3~GHz measurements are from
2002 April 29 and 8.4~GHz measurements are from 2003 February 9}
\end{deluxetable}

\begin{deluxetable}{ccccc}
\tabletypesize{\scriptsize}
\tablecolumns{5}
\tablewidth{0pc}
\tablecaption{Derived Physical Properties of the VLBI Sources}
\tablehead{
\colhead{Source}&\colhead{$\alpha_{2-8}$\tablenotemark{a}}&\colhead{$T_{\rm 
B}$\tablenotemark{b}}&
\colhead{$P$ (8.4 GHz)}&\colhead{$P/P({\rm Cas A})$} \\
\colhead{}&\colhead{}&\colhead{(K)}&\colhead{(W Hz$^{-1}$)}&\colhead{} \\}
\startdata
A0&$>2.58\pm 0.26$&$> 1.4 \times 10^{8}$&$6.5\times 10^{20}$&990 \\
A1&$-0.88\pm 0.14$&$> 2.6 \times 10^{7}$&$1.2\times 10^{20}$&180 \\
A2&$+0.39\pm 0.14$&$> 2.9 \times 10^{7}$&$1.3\times 10^{20}$&200 \\
A3&$+0.20\pm 0.20$&$> 1.7 \times 10^{7}$&$7.8\times 10^{19}$&120 \\
A4&$+0.10\pm 0.19$&$> 1.7 \times 10^{7}$&$8.0\times 10^{19}$&120 \\
D (2003 Feb 9)&\nodata&$> 9.5 \times 10^{7}$&$4.3\times 10^{20}$&680 \\
D (2003 May 1)&$+0.38\pm 0.13$&$4.3 \times 10^{7}$&$6.7\times 10^{20}$&1040 \\
\enddata
\tablenotetext{a}{Spectral indices between 2.3 and 8.4~GHz for all components
within Source A are computed using the lowest noise data from 2002 April 
(2.2~GHz) and 2003 February (8.4~GHz), since there is no evidence of
variability in these components.  Source D appears to have varied, so only
the simultaneous data from 2003 May 1 were used for computing its spectrum.}
\tablenotetext{b}{Brightness temperatures for A1, A2, A3, and A4 
are based on 8.4~GHz measurements from 2003 February; two values 
are given for Source D, for two epochs.}
\label{tab:vlba-tb}
\end{deluxetable}

\begin{deluxetable}{ccccc}
\tabletypesize{\scriptsize}
\tablecolumns{5}
\tablewidth{0pc}
\tablecaption{VLBI Flux Densities at Epoch 2003 May 01}
\tablehead{
\colhead{Source}&\colhead{$S({\rm 2.3\ GHz})$}&\colhead{$S({\rm 5.0\ GHz})$}&
\colhead{$S({\rm 8.4\ GHz})$}& \colhead{$S({\rm 15\ GHz})$} \\
\colhead{}&\colhead{(mJy)}&\colhead{(mJy)}&\colhead{(mJy)}& \colhead{(mJy)} \\}
\startdata
A0&$<0.49$&$1.08\pm 0.16$&$2.72\pm 0.19$&$2.61\pm 0.38$ \\
A1&$2.10\pm 0.26$&$0.77\pm 0.15$&$0.41\pm 0.14$&$<0.46$ \\
A2&$0.71\pm 0.17$&$0.94\pm 0.16$&$0.45\pm 0.14$&$<0.46$ \\
A3&$<0.49$&$0.45\pm 0.14$&$<0.41$&$<0.46$ \\
A4&$<0.49$&$0.50\pm 0.14$&$0.27\pm 0.14$&$<0.46$ \\
D&$2.04\pm 0.26$&$1.78\pm 0.22$&$3.36\pm 0.36$&$3.45\pm 0.54$ \\
\enddata
\label{tab:a0vlbi}
\end{deluxetable}

\begin{deluxetable}{ccccccc}
\tabletypesize{\scriptsize}
\tablecolumns{7}
\tablewidth{0pc}
\tablecaption{Individual-Epoch Flux Densities of VLA Sources}
\tablehead{
\colhead{Date}&\colhead{Config.}&\colhead{A}&\colhead{B1}&\colhead{C}&
\colhead{C$'$}&\colhead{D} \\
\colhead{}&\colhead{}& \colhead{(mJy)}& \colhead{(mJy)}&
\colhead{(mJy)}& \colhead{(mJy)}& \colhead{(mJy)} \\}
\startdata
\cutinhead{4.9~GHz }
2003 May 01 & VLBA  & $3.74\pm 0.34$ &$<0.39$ &$<0.39$ &$<0.39$ &$1.78\pm 0.16$ 
\\
2002 Apr 29 & A     &$101.7\pm 5.1$&$10.95\pm 0.59$&$4.50\pm 0.49$&$3.69\pm 
0.32$&
  $1.54\pm 0.13$ \\
2000 Oct 24 & A  &$103.7\pm 5.2$&$11.90\pm 0.60$&$7.18\pm 0.37$&$4.56\pm 0.23$&
$2.15\pm 0.11$ \\
1994 May 16 & BnA & $107.2\pm 5.4$& $14.54\pm 0.73$&$7.23\pm 0.37$&
$4.05\pm 0.21$& $1.42\pm 0.07$ \\
1993 May 07 & B & $117.6\pm 5.9$& $20.26\pm 1.03$& $11.02\pm 0.59$& $6.46\pm 
0.37$&
$1.11\pm 0.10$ \\
1993 Mar 13 & B & $113.5\pm 5.7$& $18.82\pm 0.94$& $10.74\pm 0.54$& $5.38\pm 
0.28$&
$1.40\pm 0.08$ \\
1993 Jan 28 & BnA & $103.8\pm 5.2$& $15.16\pm 0.76$& $7.34\pm 0.38$&
$4.65\pm 0.24$& $1.29\pm 0.07$ \\
1991 Jul 05 & A &$99.5\pm 5.0$& $10.86\pm 0.55$& $4.08\pm 0.24$& $4.17\pm 0.23$&
$1.17\pm 0.07$ \\
1990 May 09 & A & $99.1\pm 5.0$& $10.91\pm 0.55$& $5.15\pm 0.30$& $3.36\pm 
0.19$&
$1.53\pm 0.08$ \\
1989 Jan 14 & A & $99.6\pm 5.0$& $10.41\pm 0.56$& $3.29\pm 0.38$&$2.78\pm 0.26$&
$1.31\pm 0.11$ \\
\cutinhead{8.4~GHz }
2003 May 01 & VLBA  & $4.05\pm 0.41$ &$<0.41$ &$<0.41$ &$<0.41$ &$3.36\pm 0.36$ 
\\
2003 Feb 09 & VLBA & $5.29\pm 0.33$ &\nodata &\nodata &\nodata & $2.05\pm 0.12$ 
\\
2002 Apr 29 & A     &$69.58\pm 0.35$&$5.90\pm 0.30$&$0.66\pm 0.07$&$1.39\pm 
0.07$&
  $2.23\pm 0.11$ \\
2001 Mar 19 & B & $82.93\pm 4.15$&$12.05\pm 0.60$&$6.56\pm 0.33$& $4.62\pm 
0.23$&
$2.52\pm 0.13$ \\
1999 Oct 28 & B & $78.06\pm 3.90$&$11.06\pm 0.57$& $6.04\pm 0.36$& $3.30\pm 
0.21$&
$2.18\pm 0.13$ \\
1994 May 16 & BnA & $78.93\pm 3.95$&$7.77\pm 0.39$&$3.47\pm 0.19$&
$2.73\pm 0.15$ & $2.35\pm 0.12$ \\
1994 Feb 18 & A & $74.66\pm 3.73$ & $7.21\pm 0.37$& $1.62\pm 0.15$ & $2.82\pm 
0.18$&
 $2.54\pm 0.13$ \\
1993 May 07 & B & $84.04\pm 4.20$ & $11.67\pm 0.60$& $6.54 \pm 0.38$ & $3.85\pm 
0.23$&
$1.77\pm 0.11$ \\
1993 Mar 13 & B & $80.30\pm 4.01$ & $11.17\pm 0.56$& $5.07\pm 0.26$& $3.61\pm 
0.19$ &
$2.21\pm 0.11$ \\
1993 Jan 28 & BnA & $74.26\pm 3.71$& $7.55\pm 0.38$& $3.30\pm 0.19$ &
$2.54\pm 0.14$ & $1.91\pm 0.10 $ \\
1991 Jul 05 & A &$70.27 \pm 3.51$ & $6.21\pm 0.32$ & $1.68\pm 0.17 $ & $1.85\pm 
0.15 $&
$1.35\pm 0.08$ \\
1991 Jun 22 & A & $74.05\pm 3.70$ & $6.69\pm 0.34$ & $2.41\pm 0.20 $& $2.59\pm 
0.17$ &
$1.58\pm 0.09$ \\
1990 May 09 & A & $74.15 \pm 3.71$ & $7.22\pm 0.37$ & $1.97\pm 0.14 $& $2.64\pm 
0.16$ &
$1.71\pm 0.09 $ \\
1990 Mar 04 & A & $71.35\pm 3.57 $& $6.61\pm 0.34 $& $1.48\pm 0.15 $& $1.91\pm 
0.13 $&
$1.50\pm 0.08 $ \\
1990 Feb 24 & A & $73.83\pm 3.69$ & $6.81\pm 0.35$ & $1.12\pm 0.13 $& $1.88\pm 
0.13 $&
$1.57\pm 0.09 $ \\
\enddata
\label{tab:var}
\end{deluxetable}

%
%

\clearpage

\clearpage

\clearpage

\begin{figure}
\figurenum{6}
{\plottwo{fig6a.eps}{fig6b.eps}}
\caption{(The central region of Source A, showing the
appearance of Source A0 in the 2003 February observations.  
The left panel is a 2.3~GHz VLBA image
from 2002 April 29, using the same contour levels as figure
\ref{fig:vlba13cm-wide}:  ($-$3, 3, 6, 24) times the rms noise 
of $3.7\times 10^{-5}$ mJy~beam$^{-1}$
The right panel is an 8.4~GHz VLBA image of the same region, from 
2003 February 9. Contour levels are at ($-$3, 3, 6, 24) times the rms noise of
$5.1\times 10^{-5}$ mJy~beam$^{-1}$, and the peak flux density is
3.04~mJy~beam$^{-1}$.  The 8.4~GHz image shown here has 
been restored with a beam identical to that of the 2.3~GHz image, to
facilitate comparison.  The cross in the 2.3~GHz image (left panel) 
indicates the location of component A0.} 
\label{fig:vlba13cm-narr}
\end{figure}

\clearpage

\begin{figure}
\figurenum{8}
{\plotone{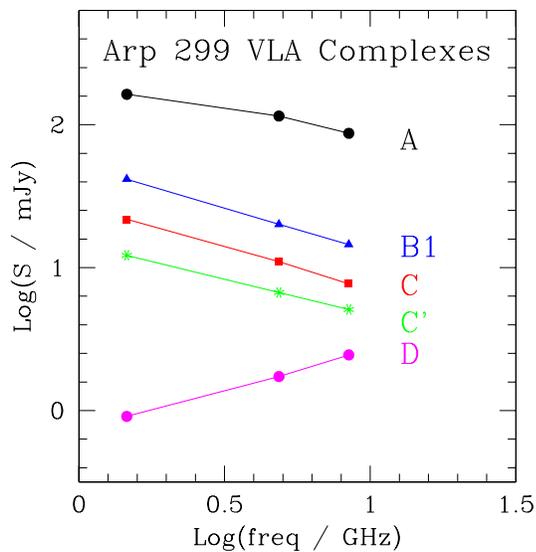}}
\caption{Radio spectra of VLA sources  
A, B1, C, C$'$, and D.
The measurements shown here are from the matched-resolution 
combined images,
all with a restoring beam of  1\dsec 21 $\times$0\dsec 88. 
Sources A, B, C, and C$'$ all have spectra 
consistent with Galactic supernova remnants.  The spectrum 
of Source D appears to peak above 8 GHz. }
\label{fig:lowres}
\end{figure}


\begin{figure}
\figurenum{9}
\epsscale{1.0}
{\plotone{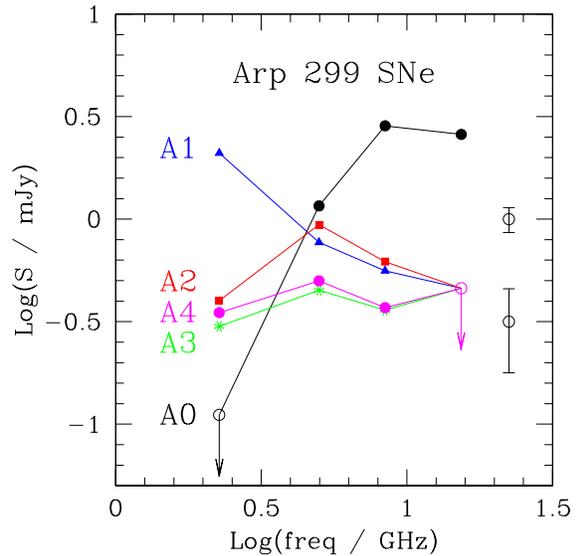}}
\caption{Radio spectra of the VLBI sources.  These measurements
are from the simultaneous observations done on 1 May 2003.
Sample error bars at log(S/mJy)$= -0.5$ and 0.0 are shown
on the right-hand side of the plot.  Error bars at
log (S/mJy)$ = 0.5$ are smaller than the sizes of the points.
Source A0 has a spectrum characteristic of a very
new supernova,  A2, A3, and A4 are all consistent with 
young supernovae in dense  environments, and A1 is suggestive
of an older supernova or supernova remnant or an AGN. }
\label{fig:vlbaspec}
\end{figure}

\begin{figure}
\figurenum{10}
\epsscale{1.0}{\plotone{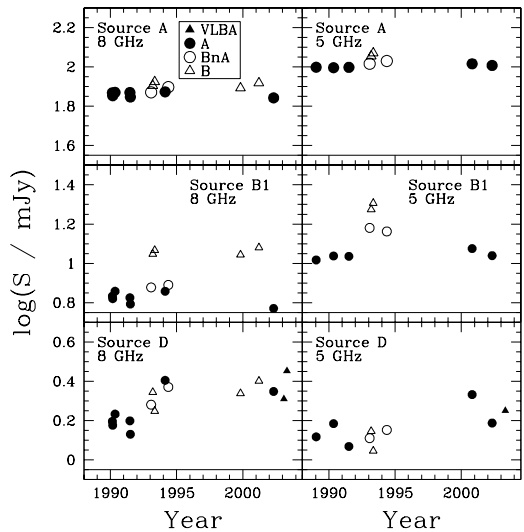}}
\caption{Flux-density history of high-resolution
observations of three radio sources in Arp 299,
as observed with resolution of $\sim 1$\arcsec\ or
better with the VLA at 4.9 and 8.4~GHz.  
Milliarcsecond measurements with the VLBA at 5.0 and 8.4~GHz also
are included for Source D, which appears completely
unresolved at these frequencies.  The different
VLA configurations and the VLBA are indicated
by different symbols, and typical flux-density
errors are approximately the same size as the symbols.
Note that Sources A and B1 show increased flux
densities as the resolution is degraded, but appear
to show no significant variability over time for a
given VLA configuration.  The unresolved Source D,
in contrast, shows significant time variability of its
flux density, and no systematic trends for higher
flux density at poorer resolution. }
\label{fig:vlavar}
\end{figure}

\end{document}